# The Unit Cell Reconstruction and Related Thermal Activation Process within Coherent Twin Boundary Migration in Magnesium

Xiao-Zhi Tang*,[1], Qun Zu[2], and Ya-Fang Guo*,[1]

[1] *Institute of Engineering Mechanics, Beijing Jiaotong University, Beijing 100044, China*
[2] *School of Mechanical Engineering, Hebei University of Technology, Tianjin 300401, China*

**Abstract** By analyzing the interface defect loop nucleation and the interface disconnection expansion in dynamic simulations, the elementary migration process of coherent twin boundary of magnesium is identified to be independent unit cell reconstruction. The atomistic pathways of the unit cell reconstruction prove their collective behavior as a stochastic response to thermal fluctuation at a stressed state, and also the onset mechanism of interface disconnection gliding: predominant pure-shuffle basal-prismatic transformation along with atomistic shear movements. The athermal shear strength, the migration barrier, the critical length of disconnection dipole and other parameters characterizing the thermal activation process are reported.

**Keywords:** Coherent Twin Boundary, Interface Disconnection Gliding, Basal-prismatic Transformation, Potential Energy Surface, Migration Barrier, Athermal Shear Strength

**Main manuscript**

1. Introduction

Based on profuse investigations on the $\{10\bar{1}2\}$ twin (known as tension twin) in hexagonal close-packed metals [1], recently researchers pointed out with substantial evidence and rigorous analysis, that $\{10\bar{1}2\}$ twin is shuffling-controlled at the nucleation stage [2, 3]. After the nucleation stage, the migration of CTB is a glide-shuffle mechanism which involves both shear and shuffle [1]. The interface disconnection with the same topological character of twinning dislocation on $\{10\bar{1}2\}$ CTB is responsible for this glide-shuffle mechanism [1]. The interface disconnection on $\{10\bar{1}2\}$ CTB has been widely confirmed in transmission electron microscopy (TEM) observations [4-8], and the mobility of such disconnection is supported by atomistic simulations [9-14], which directly contributes to twin boundary migration. Well, this is not the whole story of $\{10\bar{1}2\}$ CTB migration. The widely accepted convention is, twinning dislocation strongly interacts with precipitates so that twins are pinned down, and in this way hardening is induced [15, 16]. But the interface disconnection on $\{10\bar{1}2\}$ CTB does not. In magnesium-based alloys, such precipitation hardening is surprisingly weak: accompanied by interface disconnection gliding, tension twin engulves precipitates or even bypasses them [17-19]. Also in pure hexagonal close-packed (HCP) metals, $\{10\bar{1}2\}$ CTB deviates largely from the twinning plane [20-24]. In these deviated boundaries, basal/prismatic (BP) interface coexists with CTB confirmed by many TEM observation [6, 7, 20, 25-28]. In atomistic simulations CTB and BP interface can transform into each other [11, 13, 29]. Specifically, BP interface is supposed to be strongly associated with atomic shuffling mechanism [3, 30, 31]. Therefore, researchers began to question the conventional glide-shuffle mechanism of the migration of $\{10\bar{1}2\}$ CTB [32-36]. Some believe that it is solely accomplished by atomic shuffling [35]. The atomic shuffling mechanism in the case of $\{10\bar{1}2\}$ twin particularly refers to basal-prismatic transformation [1, 30, 37] of HCP lattice, which leads directly to unit cell reconstruction (UCR) [20, 31, 33] of tension twin. Until now, a complete description and a unified theory for $\{10\bar{1}2\}$ CTB migration at different spatio-temporal scales are still not presented. By analyzing the migration process in dynamic simulations, this study examined the connection between UCR mechanism and interface disconnection gliding. The uncovered connection is successfully testified by stress-dependent migration pathways of CTB on its potential energy surface (PES).

## 2. Simulation Methodology

The CTB migration in magnesium was investigated at 10 K by Molecular Dynamics (MD) and at 0 K by Molecular Statics (MS), adopting the EAM potential developed by Liu *et al* [38]. The simulation system contains $6.03 \times 10^5$ atoms, and the CTB could be considered infinite large since Periodic Boundary Condition (PBC) was applied along two in-plane directions. Therefore no surface effect exists. To find out the effects of simulation technique and eliminate them in the defect analysis, two ways of loading were adopted: displacing one of the outermost fixed atom layers to apply shear stress, and displacing all free atoms (sandwiched between the two fixed atom layers) to apply shear strain to the whole system. The strain rate in both loading ways is set to $3 \times 10^8 \text{ s}^{-1}$. We are not aiming to reproduce the CTB behavior at experimental strain rates and room temperature. Contrarily, extremely high strain rate in MD and low temperatures applied here are quite necessary to reveal the existence of different evolution pathways in the phase space associated with CTB structures, so that we can analyze them to find the connections. By the two loading ways mentioned above (stress-controlled, strain-controlled) and two simulation techniques (MD, MS), representative stress-strain curves are plotted in Fig. 1.

## 3. Results and Discussion

### 3.1 The elementary migration process

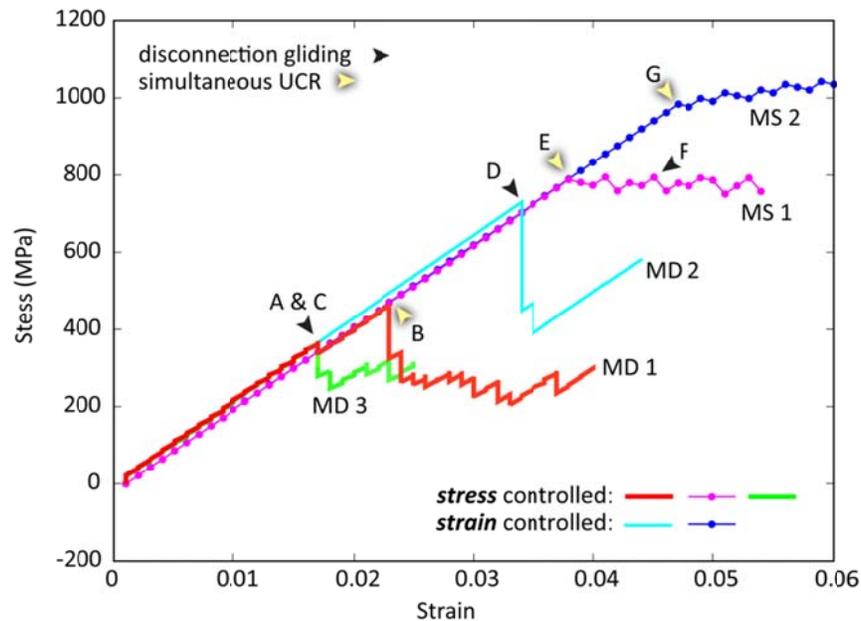

Fig. 1 The stick-slip behaviors [39, 40] of CTB reflected in shear stress fluctuation.

The plastic deformation was initiated by Disconnection Gliding (DG) in all MD simulations (point A, C, D) and by what we named as Simultaneous Transformation (ST, refers to migration in entirety) in all MS simulations (point E, G). Only in stress-controlled simulations we could see both mechanisms (point B in MD 1, F in MS 1). In MD simulations, thermal fluctuations benefit the DG mechanism for the reason presented in the next section. The ST needs cooperative thermal activations on the whole CTB which is a low-probability event. Therefore it is not dominant in MD simulations (for example no ST in MD 3). In MS simulations, there is no thermal energy, so the elementary migration process [41, 42] is purely stress/strain-driven. A defect-free CTB at 0 K is ideal for migration in entirety [43]. Therefore the ST has priority in MS simulations. The DG still exists because after the system is plastically deformed, the cooperative nucleation of elementary processes couldn't be always perfect, and the stress/strain distribution is more likely to be inhomogeneous. Stress-controlled model actually applies larger local shear strain by introducing greater relative parallel displacement between $\{10\bar{1}2\}$ atom layers than strain-controlled model does, therefore the yield stresses are lower in stress-controlled models for both MD and MS simulations. Furthermore, the variety of migration mechanisms arises from the inherent randomness of the atom-position iterations in stress-controlled model. The strain-controlled model updates atom positions artificially, so the randomness declines. Observing two mechanisms in one simulation (MD 1 and MS 1) promises inherent differences between them and thus promises comparison. The details of the two migration mechanisms are in the following.

In elastic stage in MDs, by Common Neighbor Analysis (CNA) [44], a defect is observed to appear randomly on the CTB. Fig. 2(a) shows them by coloring atoms according to their y-coordinates. The defect is named as an embryo (Fig. 2(f)). Generally the embryo lasts less than a femtosecond and disappears. If a certain area is full of embryos and the embryos inside evolve further, this part of CTB migrates and interface disconnections are left (Fig. 2(b) to (c)). All the five disconnection loops in Fig. 2(c) have mobility. From Fig. 2(c) to (e), parts of the loops annihilate. When they are all annihilated by each other, the migration process finishes (after Fig. 2(e), not shown). On the x-y cross section of the CTB, the disconnection has the same topological character of $\langle 10\bar{1}\bar{1}\rangle$ twinning dislocation (TD) in TEM observations [1, 45]. The migration process is also the same as described in previous atomistic simulations [9, 13, 46]. Accompanied by increasing stress, new embryos nucleate on the migrated CTB (Fig. 2(e)). The migration mechanism described above is DG. In the other mechanism (ST in Fig. 3), migration in entirety requires that the CTB is full of embryos and the subsequent transition is cooperative. Lattice transformations took place uniformly across the CTB, without any disconnections throughout the process. Migration in this way definitely involves no dislocation mechanism. Embryo, as small as a single lattice, is a precursor to the transformation of HCP lattice, literally the UCR. Therefore the elementary migration process in both mechanisms is confirmed to be UCR. This is the first evidence used for our interpretation of CTB migration.

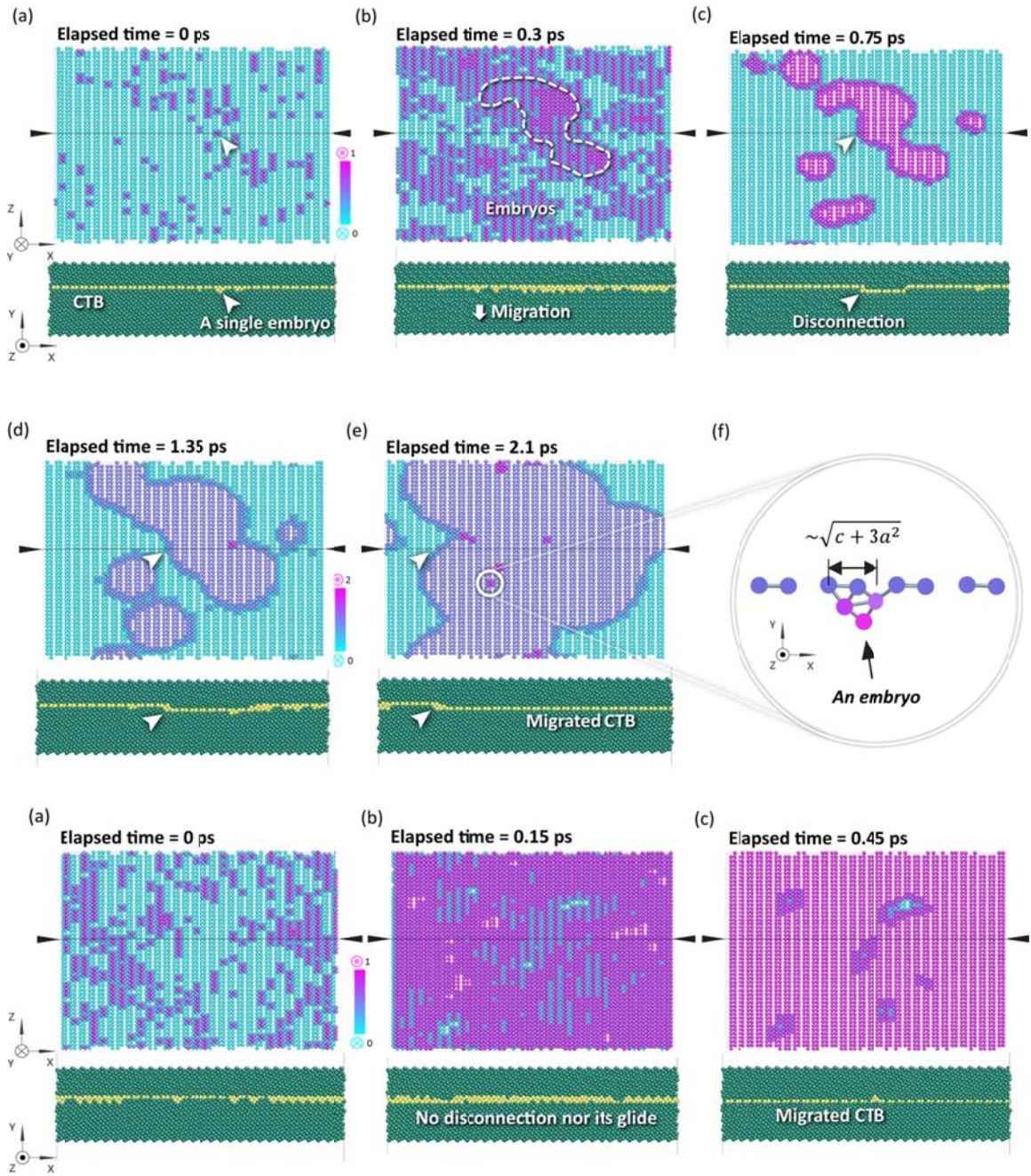

Fig. 2 Snapshots of DG mechanism (a-e) and ST mechanism (g-i) captured in MD 1, taken at strain of 0.017 (point A) and at strain of 0.024 (point B) separately. The total elapsed time is 2.1 ps for DG and 0.45 ps for ST. X-axis is along $\langle 10\bar{1}\bar{1}\rangle$ direction and z-axis is along $\langle 11\bar{2}0\rangle$ direction. The number of color bar indicates the neighboring position sequence of CTB.

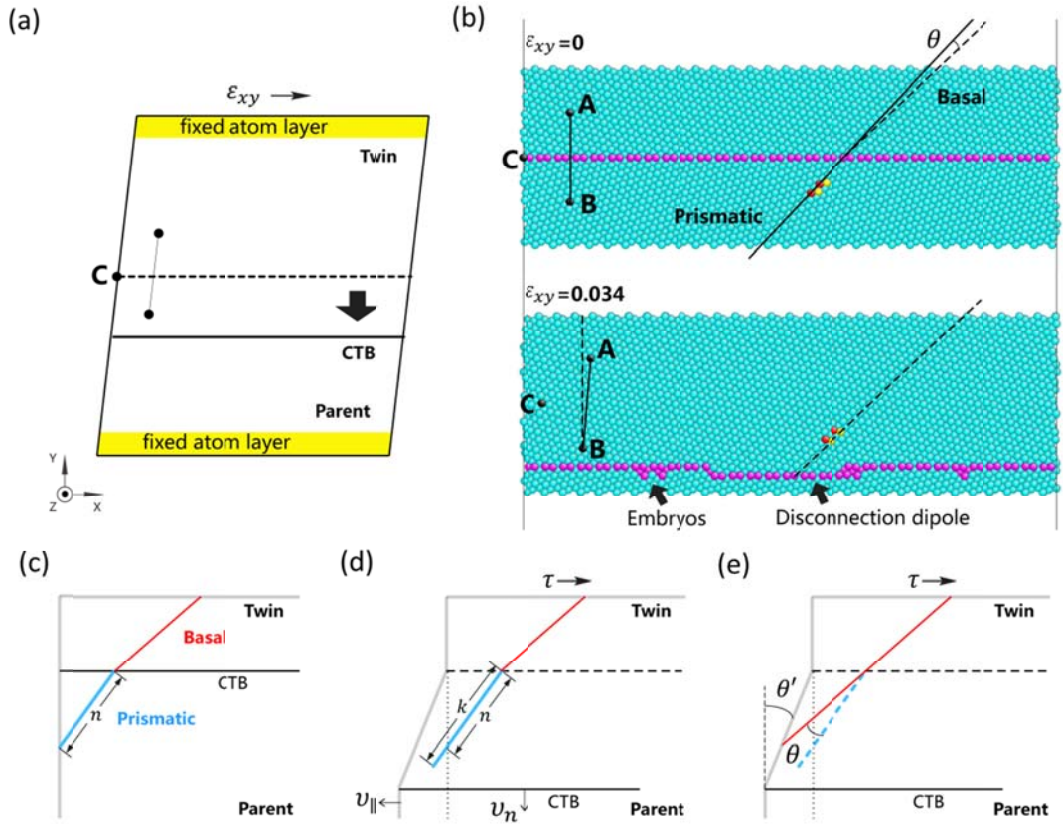

Fig. 3 (a-b) The basal-prismatic transformation created by interface disconnection gliding. (c-e) Coupling CTB migration to shear deformation by two components: lengthening (atomistic shuffling) and rotating (accumulated shear movements). Note that the included angle $\theta$ is larger than the actual value for visual clearance.

We dug deeper on the properties of UCR in CTB migration, for the evidence of basal-prismatic transformation, and its connections with interface disconnection gliding. In Fig. 3, strain-controlled model is illustrated in (a), where original position of CTB is plotted by the dashed line. Two tagged atoms (A and B) are shown as an index of shear deformation, and atom C is the index of relative translation of the grains parallel to the CTB. In Fig. 3(b), before migration the basal plane of the twin is indexed by dashed line and the prismatic plane is indexed by solid line. They have a included angle $\theta$ which is equal to $\pi/4 - \tan^{-1}(c/\sqrt{3}a)$, essentially caused by c/a ratio not equal to $\sqrt{3}$. Red and yellow atoms are originally belong to two adjacent prismatic planes. After interface disconnections swept these four atoms (CTB migrates over), they constitute the basal planes separately. This clearly and undoubtedly demonstrates the basal-prismatic transformation: the atoms of prismatic plane in the parent exactly are the atoms of basal plane in the twin, no more, no less. The basal-prismatic transformation itself does not require shear but only the atomic shuffling [30, 31, 37]. So, where is the shear which indeed exists in the case of $\{10\bar{1}2\}$ twin? In the conventional diagram of grain boundary migration coupled with shear deformation, this question can be explained. In Fig. 3 (c-e), the untransformed prismatic plane in the parent has a length of $n$. After transformed into basal plane with a length of $k$, it has to rotate by angle $\theta$ to align with the basal plane in the twin. The twinning shear $\theta'$ in Fig. 3 (e) is brought both by the lengthening and rotating. But the shear movement of atoms in HCP lattices is actually not that large, because the lengthening is a pure-shuffle mechanism transforming the basal into the prismatic, and vice versa. Twinning shear is mainly contributed by atomic shuffling, and the main effect of disconnection gliding is a pure-shuffle transformation. This is the second evidence used for our interpretation of CTB migration.

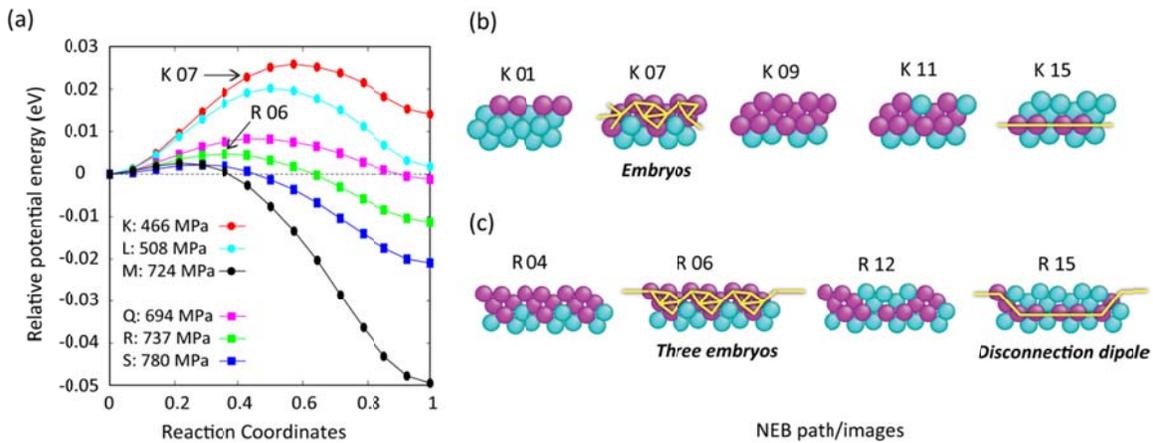

Fig. 4 (a) Shear stress-dependent minimum energy paths (MEP) separately of simultaneous transformation (K, L, M) and disconnection gliding (Q, R, S). (b)(c) Atomic images of transition states.

To uncover the transition processes of ST and DG, calculate both the activation parameters and athermal shear strength, we explored the pathways of migration on the zero-temperature potential energy surface (PES) using the climbing image nudged elastic band (CINEB) method [47]. Defects along $\langle 10\bar{1}\bar{1} \rangle$ direction are prohibited by reducing the system dimension to one lattice along $\langle 11\bar{2}0 \rangle$ direction. The final states of both mechanisms at different stresses were detected by autonomous basin climbing (ABC) method [48-52]. The NEB pathways and corresponding atomic images of a two-element disconnection dipole and a CTB segment which is also two-element long are in Fig. 4(a)-(c). Here one element represents a smallest size CTB in unit of lattice spacing. The actual dimension of the sample along x-axis is way larger than three elements, but all the presented values are obtained only from the structures shown in the figure. In ST, embryo is a transition state (K07 in Fig. 4(b)) before the CTB is fully activated. The atomistic saddle-point state (K09) has a more symmetric GB structure different from embryos. The activation potential energy is 0.026 eV. In DG, embryo is the saddle-point state. Across the saddle point, the two embryos on the edge of activated CTB evolve into disconnections and the embryos inside accomplish the transition (state R06 to R15 in Fig. 4(c)). NEB paths prove that the interface disconnection is inherently evolved from embryo. This is the third evidence used for our interpretation of CTB migration.

As said, an embryo is a precursor to UCR. Tracing back to Fig. 2, disconnection loops could be any size and shape (in a certain range), and the disconnection gliding can be seen as the transformation between two nucleation events of different-size loops expanding on time scale. The nucleation of loops requires nothing but embryos, so line defect gliding on the interface is constituted by successive UCRs. This is in accord with our first and third evidences mentioned above. What does UCR in the case of $\{10\bar{1}2\}$ CTB migration really mean? Although the basal-prismatic transformation already reconstructs HCP lattice, there is still shear in the UCR, to make the orientational relationship 86°, not 90°. This is clearly reflected in Fig. 3 (c-e). In this way, UCR in the case of $\{10\bar{1}2\}$ CTB migration has an extended meaning than previously proposed [ ]. All three evidences are indicating the glide-shuffle mechanism applied on $\{10\bar{1}2\}$ CTB is slightly different from the conventional one. Here the atomic shuffling is predominant, and responsible for the more effective part of an interface disconnection, while not the shear.

## 3.2 The thermal activation process

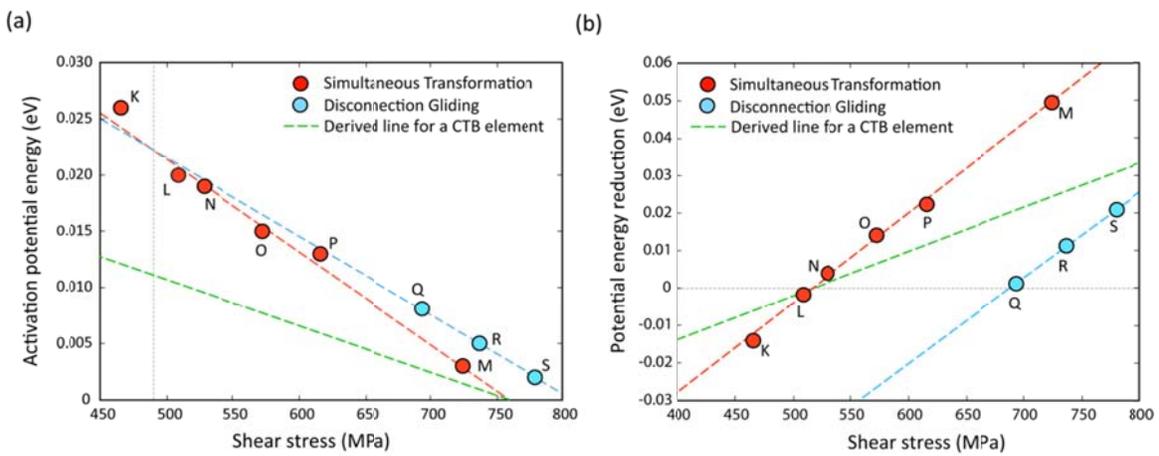

Fig. 5 (a) Activation potential energy. (b) Potential energy reduction.

The activation potential energy as a function of shear stress is plotted in Fig. 5(a). Liner fittings enable us to describe the stress dependency of the two migration mechanism. To be noted the calculated barrier is for a system with dimension of two CTB elements, so the intrinsic activation parameters can be gotten by the derived line for one CTB element: 1), The athermal shear strength $\tau_a$ is 759 MPa. 2), The activation volume of elementary migration process $\Omega_e$ is -4.22E-5 eV/MPa. 3), The activation energy of elementary migration process at stress free condition $\Delta U_0$ is 0.032 eV. 4), The migration barrier at stress free condition $\gamma_\infty$ is 21.14 mJ/m². Clearly, the existence of disconnection dipole raises the barrier at certain stresses. The increment per unit length of disconnection $\Delta U^{dd}$ reflects the slope difference of the two fitted lines and is stress-dependent. Apparently, if the final state of transition is not in a lower energy state, satisfying the barrier does not realize the transition. That means the migrated CTB is unstable unless local shear stress make it a lower energy state. The potential energy reductions as a function of shear stress are plotted in Fig. 4(e). For simultaneous UCR, the reduction in unit of energy density is also linearly stress-dependent. Therefore the effective ranges of four physical factors for thermal migration are confirmed: 1), Stress: 518 MPa ($\tau_f$) < $\tau$ < 759 MPa ($\tau_a$). 2), The corresponding energy reduction: 0 mJ/m² < $\gamma^*$ < 20.78 mJ/m². 3), The corresponding migration barrier: 6.60 mJ/m² > $\gamma$ > 0 mJ/m². Considering the temperature and strain rate effect, all the calculated values basically ensure the self-consistency within our model. Due to the unavoidable calculation inaccuracies, they are not preferred in precise comparisons, but valid for qualitative analysis. Apparently the existence of disconnection dipole makes the energy reduction less effective, the critical length of a dipole could be analyzed in thermodynamics (the pure mechanical factors are not considered). At stress of 688 MPa ($\tau_f$ for disconnection dipole of two two-element CTB), energy reduction of simultaneous UCR is 0.041 eV. Since the slopes of the two mechanisms (red and blue dash lines in Fig. 4(e)) are quite the same, the 0.041 eV is supposed to be stress-independent, as well as the $\gamma^*_{dd}$=0.041 eV/$2l_0$=0.064 eV/nm. At a certain stress, the critical length of disconnection dipole $l_c^{dd}$ in the unit of CTB element is

$$\frac{\gamma^*_{dd}*2l_0}{\gamma^**A_0} + 1 = \frac{0.041}{1.21\times10^{-4}\tau-0.06} + 1 \tag{1}$$

According to Eq. (1), 835 MPa is the athermal stress for disconnection dipole nucleation ($\tau_a^{dd}$), and the inhomogeneous distribution on CTB is still indispensable. At stress of 759 MPa ($\tau_a$), the $l_c^{dd}$ is 3 CTB elements. At stress of 518 MPa ($\tau_f$), the $l_c^{dd}$ is 17 CTB elements (12.90 nm), and the activation potential energy is calculated to be around 0.16 eV (2.5 meV/atom). Definitely the thermal energy $k_B T$ at room temperature (26 meV) is enough large. So theoretically $\tau_f$ becomes the lower limit of the migration stress window for bulk magnesium, where the characteristic length of CTB is usually at micrometer scale. As we can see, potential energy barrier is generally small and wouldn't be a problem. Stress level, critical length of disconnection dipole, and inhomogeneous distributions of physical quantities are the three requirements for shuffle-induced migration. Once again, all the conclusions here do not involve the disconnections along $\langle 10\bar{1}\bar{1} \rangle$ direction. In practice, disconnection gliding has to be analyzed as a loop for a more comprehensive understanding.

It is clear that a single embryo corresponds to a transition state in the migration process of a CTB element. Energetically and mechanically it is hard to finish migration for just one CTB element. For a certain area, thermal fluctuations have to activate all the elementary migration processes inside (indexed by embryos) to make cooperative transition physically possible, and local shear stress has to lower the energy state of migrated CTB with the given ratio of (disconnection length)/(the area) to make the cooperative transition energetically possible. If all the physical conditions are satisfied, statistically we would always witness disconnection gliding when $\{10\bar{1}2\}$ CTB migrates. The thermal energy at 298 K (26 meV) is enough high for the elementary migration process (8 meV/atom). However the CTB deviated from original position is not an equilibrium state without shear stress (at least ~510 MPa is needed according to Fig. 5(b)). Therefore $\{10\bar{1}2\}$ CTB is thermally stable at room temperature or even higher. The shear stress along $\langle 10\bar{1}\bar{1} \rangle$ produces asymmetry of the barriers rendering directional bias to the embryo nucleation, and also makes the MEP favorable by system evolution. At stress below athermal strength, the Brownian motion of CTB at high temperature [39, 40, 53] is expected. At stress above athermal strength, CTB migration turns into pure stress-driven process. The nature of nucleation bias of embryos lies in the lattice transformation from parent to twin. Two misfit strain components are produced by the shear strain/stress loading [35]. Therefore, to accommodate the local misfit strains, the direction of CTB migration is certain and reflects on the bias of embryo nucleation.

Generally the stick-slip dynamics of CTB is characterized by a saw-tooth time/strain/grain-translation dependence of the stress and the motion in a stop-and-go manner: stress fluctuates periodically with equal peak values and CTB migrates to the nearest neighboring position in each slip event. This is not reflected well in Fig. 1. High strain rate and low temperature were applied to realize simultaneous UCR mechanism, but the accumulated deformation is also released in an abrupt way that the peak values varies drastically and migration spans several neighboring positions. At a lower strain rate of $3 \times 10^7$ s$^{-1}$, standard stop-and-go manner is recovered at 298 K. The peak value is 38 MPa, significantly smaller than the other temperatures (155 MPa at 150 K and 590 MPa at 10 K). The temperature sensitivity suggests that free energy calculations should be carried out for addressing accurate thermo-mechanical characteristics. Existing theory of boundary migration [39] describes the rate dependency by treating the interface defect nucleation as governed by its activation energy. While it does not work well for $\{10\bar{1}2\}$ CTB migration at room temperature, for what NEB calculations above already proves: 1), Whether activation energy $\Delta G$ or migration barrier $\gamma$ is barely larger than $k_B T$. 2), What triggers the migration is not a reduced barrier, but a lower final-state energy which is quite sensitive to stress/strain. Yet analytical approaches independent of activation energy needs to be applied here.

## 4. Conclusions

In conclusion, by investigations on the kinetics and MEPs of $\{10\bar{1}2\}$ CTB migration, we found the successive unit cell reconstruction mainly composed by atomic shuffling is the onset mechanism of the interface disconnection gliding, and the seemingly stochastic behavior of the elementary migration processes reflects the stress and temperature effects on activation parameters. Thus we describe the $\{10\bar{1}2\}$ CTB migration as following outlines:

1), Pure-shuffle basal-prismatic transformation along with shear movements intrinsically induced by c/a ratio constitutes the interface disconnection gliding.

2), Thermal activation of disconnection loop is the collective behavior of the elementary migration processes.

3), Shear stress makes unit cell reconstruction physically realizable not by reducing the energy barrier, but lowering the final energy state, and meantime, shorten the critical length of a disconnection dipole along $\langle 10\bar{1}\bar{1} \rangle$ direction.

Overall, the intrinsic shuffling mechanism and the efficacy of mobile interface disconnection are both included in one framework presented in this article. Atomic simulations at realistic laboratory strain rates and free energy calculations can be available soon, providing more conclusive evidence.

**Acknowledgement**

This work was supported by National Natural Science Foundation of China (No. 11372032 and No. 11602015). XZT sincerely appreciates the hospitality of the Nuclear Science and Engineering Department at MIT for an academic visit.

**References**

1. Hirth, J., J. Wang, and C. Tomé, *Disconnections and other defects associated with twin interfaces.* Progress in Materials Science, 2016. **83**: p. 417-471.

2. Ishii, A., J. Li, and S. Ogata, *Shuffling-controlled versus strain-controlled deformation twinning: The case for HCP Mg twin nucleation.* International Journal of Plasticity, 2016. **82**: p. 32-43.

3. Wang, J., et al., *Pure-shuffle nucleation of deformation twins in hexagonal-close-packed metals.* Materials Research Letters, 2013. **1**(3): p. 126-132.

4. Braisaz, T., et al., *Investigation of {10 1 2} twins in Zn using high-resolution electron microscopy: Interfacial defects and interactions.* Philosophical Magazine A, 1997. **75**(4): p. 1075-1095.

5. Braisaz, T., *High-resolution electron microscopy study of the (1012) twin and defects analysis in deformed polycrystalline alpha titanium.* Philosophical magazine letters, 1996. **74**(5): p. 331-338.

6. Sun, Q., et al., *Interfacial structure of {} twin tip in deformed magnesium alloy.* Scripta Materialia, 2014. **90**: p. 41-44.

7. Tu, J., et al., *Structural characterization of {10(1)overbar2} twin boundaries in cobalt.* Applied Physics Letters, 2013. **103**(5): p. 1903.

8. Tu, J., et al., *HREM investigation of twin boundary and interface defects in deformed polycrystalline cobalt.* Philosophical Magazine Letters, 2013. **93**(5): p. 292-298.

9. Serra, A. and D.J. Bacon, *Interaction of a moving {10(1)over-bar2} twin boundary with perfect dislocations and loops in a hcp metal.* Philosophical Magazine, 2010. **90**(7-8): p. 845-861.

10. Pond, R., A. Serra, and D. Bacon, *Dislocations in interfaces in the hcp metals—II. Mechanisms of defect mobility under stress.* Acta materialia, 1999. **47**(5): p. 1441-1453.

11. Wang, J., et al., *Twinning-associated boundaries in hexagonal close-packed metals.* Jom, 2014. **66**(1): p. 95-101.

12. Barrett, C.D. and H. El Kadiri, *Impact of deformation faceting on and embryonic twin nucleation in hexagonal close-packed metals.* Acta Materialia, 2014. **70**: p. 137-161.

13. Ostapovets, A. and R. Gröger, *Twinning disconnections and basal–prismatic twin boundary in magnesium.* Modelling and Simulation in Materials Science and Engineering, 2014. **22**(2): p. 025015.

14. Barrett, C.D. and H. El Kadiri, *The roles of grain boundary dislocations and disclinations in the nucleation of {10 2} twinning.* Acta Materialia, 2014. **63**: p. 1-15.

15. Nie, J.-F., *Precipitation and hardening in magnesium alloys.* Metallurgical and Materials Transactions A, 2012. **43**(11): p. 3891-3939.

16. Ardell, A., *Precipitation hardening.* Metallurgical Transactions A, 1985. **16**(12): p. 2131-2165.

17. Robson, J., N. Stanford, and M. Barnett, *Effect of precipitate shape on slip and twinning in magnesium alloys.* Acta materialia, 2011. **59**(5): p. 1945-1956.

18. Gharghouri, M., G. Weatherly, and J. Embury, *The interaction of twins and precipitates in a Mg-7.7 at.% Al alloy.* Philosophical Magazine A, 1998. **78**(5): p. 1137-1149.

19. Stanford, N. and M. Barnett, *Effect of particles on the formation of deformation twins in a magnesium-based alloy.* Materials Science and Engineering: A, 2009. **516**(1): p. 226-234.

20. Tu, J. and S. Zhang, *On the 10$\bar{1}$2 twinning growth mechanism in hexagonal close-packed metals.* Materials & Design, 2016. **96**(Supplement C): p. 143-149.

21. Zhang, X., et al., *Twin boundaries showing very large deviations from the twinning plane.* Scripta Materialia, 2012. **67**(10): p. 862-865.

22. Zhang, Y., et al., *An analytical approach and experimental confirmation of dislocation-twin boundary interactions in titanium.*

23. Li, B. and X.Y. Zhang, *Global strain generated by shuffling-dominated twinning.* Scripta Materialia, 2014. **71**: p. 45-48.

24. Zhang, X., et al., *Non-classical twinning behavior in dynamically deformed cobalt.* Materials Research Letters, 2015. **3**(3): p. 142-148.

25. Tu, J., et al., *Structural characterization of irregular-shaped twinning boundary in hexagonal close-packed metals.* Materials Characterization, 2015. **106**: p. 240-244.

26. Tu, J., et al., *Structural characterization of {102} twin tip in deformed magnesium alloy.* Materials Characterization, 2015. **110**: p. 39-43.

27. Sun, Q., et al., *Characterization of basal-prismatic interface of twin in deformed titanium by high-resolution transmission electron microscopy.* Philosophical

Magazine Letters, 2015. **95**(3): p. 145-151.

28. Lay, S. and G. Nouet, *Morphology of (01·2) twins in zinc and related interfacial defects.* Philosophical Magazine A, 1995. **72**(3): p. 603-617.

29. Zu, Q., et al., *Atomistic study of nucleation and migration of the basal/prismatic interfaces in Mg single crystals.* Acta Materialia, 2017. **130**: p. 310-318.

30. Liu, B.-Y., et al., *Twinning-like lattice reorientation without a crystallographic twinning plane.* Nat Commun, 2014. **5**.

31. Liu, B.-Y., et al., *Terrace-like morphology of the boundary created through basal-prismatic transformation in magnesium.* Scripta Materialia, 2015. **100**: p. 86-89.

32. Zhiwei, S. and L. Boyu, *THE MECHANISM OF {10 (1) over-bar2} DEFORMATION TWINNING IN MAGNESIUM.* ACTA METALLURGICA SINICA, 2016. **52**(10): p. 1267-1278.

33. Liu, B.-Y., Z.-W. Shan, and E. Ma, *Non-Dislocation Based Room Temperature Plastic Deformation Mechanism in Magnesium*, in *Magnesium Technology 2016*. 2016, Springer. p. 199-201.

34. Cayron, C., *Hard-sphere displacive model of extension twinning in magnesium.* Materials & Design, 2017. **119**(Supplement C): p. 361-375.

35. Li, B. and X. Zhang, *Twinning with zero twinning shear.* Scripta Materialia, 2016. **125**: p. 73-79.

36. Cayron, C. and R. Loge, *Evidence of new twinning modes in magnesium questioning the shear paradigm.* arXiv preprint arXiv:1707.00490, 2017.

37. Li, B. and E. Ma, *Atomic Shuffling Dominated Mechanism for Deformation Twinning in Magnesium.* Physical Review Letters, 2009. **103**(3): p. 035503.

38. Liu, X.-Y., et al., *Anisotropic surface segregation in Al-Mg alloys.* Surface Science, 1997. **373**(2–3): p. 357-370.

39. Mishin, Y., et al., *Stick-slip behavior of grain boundaries studied by accelerated molecular dynamics.* Physical Review B, 2007. **75**(22): p. 224101.

40. Ivanov, V. and Y. Mishin, *Dynamics of grain boundary motion coupled to shear deformation: an analytical model and its verification by molecular dynamics.* Physical Review B, 2008. **78**(6): p. 064106.

41. Rajabzadeh, A., et al., *Elementary mechanisms of shear-coupled grain boundary migration.* Physical review letters, 2013. **110**(26): p. 265507.

42. Combe, N., F. Mompiou, and M. Legros, *Disconnections kinks and competing modes in shear-coupled grain boundary migration.* Physical Review B, 2016. **93**(2): p. 024109.

43. Sutton, A.P., et al., *Interfaces in Crystalline Materials and Surfaces and Interfaces of Solid Materials.* Physics Today, 1996. **49**: p. 88.

44. Faken, D. and H. Jónsson, *Systematic analysis of local atomic structure combined with 3D computer graphics.* Computational Materials Science, 1994. **2**(2): p. 279-286.

45. Serra, A., D.J. Bacon, and R.C. Pond, *Comment on "Atomic Shuffling Dominated Mechanism For Deformation Twinning In Magnesium".* Physical Review Letters, 2010. **104**(2).

46. Ostapovets, A., P. Molnár, and R. Gröger. *On basal-prismatic twinning interfaces in magnesium*. in *IOP conference series: Materials Science and Engineering*. 2014. IOP Publishing.

47. Henkelman, G., B.P. Uberuaga, and H. Jónsson, *A climbing image nudged elastic band method for finding saddle points and minimum energy paths.* The Journal of Chemical Physics, 2000. **113**(22): p. 9901-9904.

48. Kushima, A., et al., *Computing the viscosity of supercooled liquids.* The Journal of chemical physics, 2009. **130**(22): p. 224504.

49. Fan, Y., A. Kushima, and B. Yildiz, *Unfaulting mechanism of trapped self-interstitial atom clusters in bcc Fe: A kinetic study based on the potential energy landscape.* Physical Review B, 2010. **81**(10): p. 104102.

50. Fan, Y., et al., *Mechanism of Void Nucleation and Growth in bcc Fe: Atomistic Simulations at Experimental Time Scales.* Physical Review Letters, 2011. **106**(12): p. 125501.

51. Fan, Y., et al., *Onset Mechanism of Strain-Rate-Induced Flow Stress Upturn.* Physical Review Letters, 2012. **109**(13).

52. Yan, X., et al., *Atomistic modeling at experimental strain rates and timescales.* Journal of Physics D: Applied Physics, 2016. **49**(49): p. 493002.

53. Zhang, H., et al., *Characterization of atomic motion governing grain boundary migration.* Physical Review B, 2006. **74**(11): p. 115404.